%% file: MAIN.tex
\documentclass[sigconf, authorversion=true]{acmart}
\usepackage{textcomp}

\usepackage{multirow}
\usepackage{tikz} 
\usepackage{mathtools}
\usepackage{calc} 
\usepackage{hyperref}
\usepackage{xfrac}
\usepackage{siunitx}

\usepackage{colortbl}
\usepackage{tabularray} 
\UseTblrLibrary{booktabs} 
\usepackage{subcaption}
\usepackage{algorithm}
\usepackage{algpseudocode} 

\AtBeginDocument{%
  }

\copyrightyear{2025}
\acmYear{2025}
\setcopyright{acmcopyright}
\setcctype{by}
\acmConference[CVMP '25]{European Conference on Visual Media Production}{December 3--4, 2025}{London, United Kingdom}
\acmBooktitle{European Conference on Visual Media Production (CVMP '25), December 3--4, 2025, London, United Kingdom}\acmDOI{10.1145/3756863.3769703}
\acmISBN{979-8-4007-2117-5/2025/12}

\begin{document}

\title{Multi Task Denoiser Training for Solving Linear Inverse Problems}

\author{Cl\'ement Bled}
\email{cbled@tcd.ie}
\orcid{0000-0001-5395-5773}
\authornotemark[1]
\affiliation{%
  \institution{Trinity College}
  \city{Dublin}
  \country{Ireland}
}

\author{Fran\c{c}ois Piti\'e}
\email{frcs@tcd.ie}
\orcid{0000-0003-4599-0549}
\affiliation{%
  \institution{Trinity College}
  \city{Dublin}
  \country{Ireland}
}
\begin{abstract}
Plug-and-Play Priors (PnP) and Regularisation by Denoising (RED) have established that image denoisers can effectively replace traditional regularisers in linear inverse problem solvers for tasks like super-resolution, demosaicing, and inpainting. It is now well established in the literature that a denoiser's residual links to the gradient of the image log prior (Miyasawa and Tweedie), enabling iterative, gradient ascent-based image generation (e.g., diffusion models), as well as new methods for solving inverse problems.  Building on this, we propose enhancing Kadkhodaie and Simoncelli’s gradient-based inverse solvers by fine-tuning the denoiser within the iterative solving process itself. Training the denoiser end-to-end across the solver framework and simultaneously across multiple tasks yields a single, versatile denoiser optimised for inverse problems. We demonstrate that even a simple baseline model fine-tuned this way achieves an average PSNR improvement of +1.34 dB across six diverse inverse problems while reducing the required iterations. Furthermore, we analyse the fine-tuned denoiser's properties, finding that its optimisation objective implicitly shifts from minimising standard denoising error (MMSE) towards approximating an ideal prior gradient specifically tailored for guiding inverse recovery.
\end{abstract}



\maketitle

\section{Introduction}

Denoising networks~\cite{bled2022assessing, zhang_beyond_2017,liu2021swin,tassano2019dvdnet,tassano2020fastdvdnet,guo2019toward, yue2023rvideformer,liang2022vrt,yan2020depth,ren2022robust,qiao2017learning} have achieved remarkable success by implicitly learning intricate image priors from massive datasets. These networks excel at removing noise while preserving fine details and structural information, suggesting that they have captured fundamental properties of natural images.

Leveraging this implicit knowledge, recent research has explored using denoisers as priors for solving inverse problems. 
%
This prior was first indirectly exposed through the ``Plug-and-Play" priors (PnP) approach by Venkatakrishnan~\cite{venkatakrishnan2013plug} by incorporating denoisers into the regularisation term of the optimisation problem. More recently, Kadkhodaie and Simoncelli~\cite{kadkhodaie_solving_2021, kadkhodaie2021stochastic} were able to make a direct connection between the denoiser and the prior $p(x)$ by exploiting Tweedie\cite{robbins1992empirical} and Miyasawa's~\cite{miyasawa1961empirical} relationship between the residual image of a blind AWGN denoiser, \( \hat{x}(y) - y \)  and the gradient of the log-likelihood of the image:
\begin{equation}
\hat{x}(y) - y = \frac{1}{\sigma^2}\ \nabla_{y}\log{p(y)},   
\label{eqn:chp6_propto}
\end{equation}
Although this connection was not explicitly made at the time, it also underpins the functionality of contemporary diffusion models~\cite{sohl2015deep, song2020denoising, dhariwal2021diffusion,nichol2021improved,wu2024jores,wu2024diffusion,lu20243d,tan2024blind}. While both PnP methods and Kadkhodaie's solvers present iterative approaches for solving linear inverse problems, the former relies on using denoisers as a regulariser within an optimisation framework, whereas the latter leverages Equation~\ref{eqn:chp6_propto}, offering a more streamlined approach.

In this paper, we propose to take this problem on its head and demonstrate that the implicit image prior captured by the denoiser can be fine-tuned through Kadkhodaie's framework to improve its versatility as a problem-solving network. To accomplish this, we embed a pre-trained denoiser within the iterative framework and fine-tune it to generalise more effectively across tasks such as image super-resolution, inpainting, and demosaicing. By retraining the denoiser, we once again transform the natural denoiser prior into a universal image solver capable of addressing a wide array of imaging challenges. 
Our key contributions are thus as follows.
\begin{itemize}
    \item We demonstrate the effectiveness of fine-tuning a denoiser within the framework of Kadkhodaie and Simoncelli for improved linear inverse problem solving.
    \item We evaluate the performance of the fine-tuned denoiser for use on various inverse problems, including super-resolution, inpainting, and demosaicing
    \item We analyse the impact of fine-tuning on the denoiser's general denoising performance.
\end{itemize}

\section{Background}
\subsection{Plug-and-Play Prior}
The concept of using non-deep-learning denoisers to tackle a variety of imaging problems was first introduced in 2013 by Venkatakrishnan, Bouman, and Wohlberg~\cite{venkatakrishnan2013plug} in their \textit{Plug-and-Play Prior} (PnP) framework. By treating these sophisticated denoisers as priors within an optimisation setup, they demonstrated how existing denoising algorithms could be repurposed to solve inverse problems. In classical inverse problems, the objective is to estimate the unknown image \( \hat{x} \) by solving:
\[
\hat{x} = \arg \min_x \{ l(y;x) + \beta s(x) \},
\]
where \(l(y;x)\) represents the data-fidelity term, \(s(x)\) is a regulariser encoding prior knowledge of natural images and \(\beta\) controls the weight of the prior. Plug-and-Play (PnP) methods use variable splitting to separate data-fidelity from prior terms. Instead of directly minimizing both together, they introduce an auxiliary slack variable, traditionally updated via the Alternating Direction Method of Multipliers (ADMM)~\cite{boyd2011distributed}. In PnP, the denoiser acts as a proxy for this slack variable, serving as an implicit prior that guides the solution without requiring an explicit analytic prior function.

This innovation demonstrated the potential of state-of-the-art denoisers for inverse problems such as X-ray, MRI~\cite{venkatakrishnan2013plug} reconstruction and super-resolution~\cite{brifman2016turning}. This framework has since then been simplified~\cite{romano2017little} and extended to DNN denoisers~\cite{zhang_learning_2017,zhang2021plug}. These advances demonstrated significant improvements across tasks like deblurring, super-resolution, and demosaicing, solidifying PnP’s versatility.

\subsection{Solving Inverse Problems Using Denoisers Implicit Priors}

\begin{figure}
    \centering
    \includegraphics[width=\linewidth]{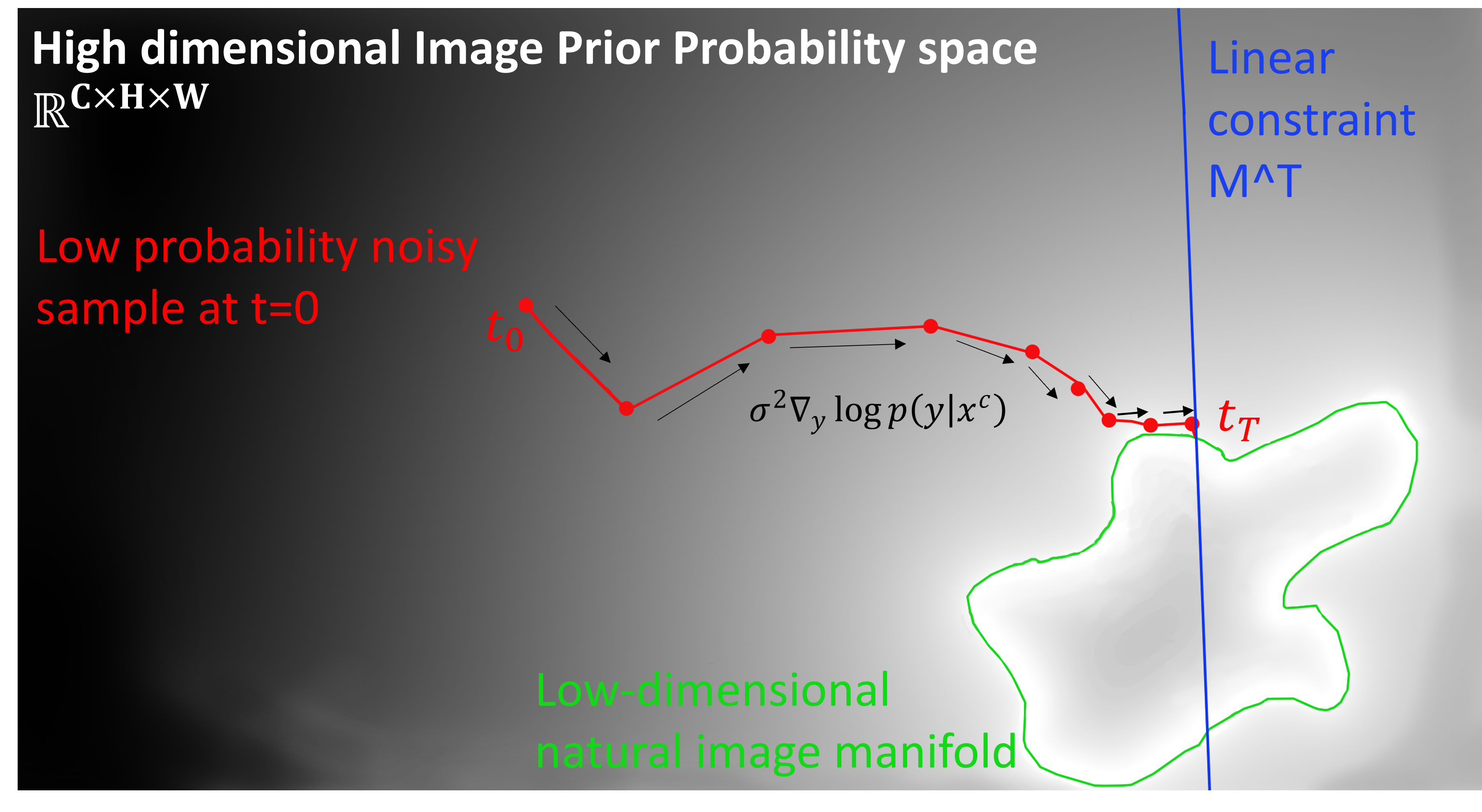}
    \caption{Illustration of solving a linear inverse problem in image probability space \( \mathbb{R}^{C \times H \times W} \). Starting from a low-probability sample \( y_0 \) (red point, left), the solution is iteratively refined via gradient ascent on the log posterior \( \nabla_y \log p(y \mid x^c) \), guided by the denoiser's residual. The red trajectory represents the evolution of the sample toward higher-probability regions, constrained to remain within the linear subspace defined by the data fidelity constraint \( M^\top y = x^c \) (blue line). The solution converges near the natural image manifold (outlined in green), which captures high-probability images under the learned prior.}
    \label{fig:log_prob_space_demo}
    \Description{The figure shows a grayscale gradient background that transitions from black on the left to a white irregular shape on the right, where the white regions represent high-probability natural images. A green outline marks this shape as the natural image manifold. Starting on the left in a dark, low-probability region, a red dot traces a zig-zag path toward a vertical blue line that represents the measurement constraint, ending near the white manifold. The path depicts the iterative refinement of a noisy initial sample toward a realistic image while satisfying the constraint. } 
\end{figure}

In contrast to PnP's indirect approach, Kadkhodaie and Simoncelli introduced a new method that allows the denoiser image prior to be used more directly, enhancing the interpretability and effectiveness of inverse problem solutions. This method relies on the rediscovered Tweedie-Miyasawa~\cite{robbins1992empirical, miyasawa1961empirical, efron2011tweedie} result, expressed in Equation~\ref{eqn:chp6_propto}, which provides a mathematical connection between the denoising residual and the gradient of the log-probability. While this interpretation is mathematically equivalent to that used in diffusion models~\cite{daras2024survey}, we continue with Kadkhodaie’s formulation for its simplicity and clarity in illustrating the iterative solving framework.




Leveraging this result, Kadkhodaie and Simoncelli created a framework to approach the low-dimensional manifold of real images, which maximises \(p(x)\), within the high-dimensional space of possible images $\mathbb{R}^N$ (with $N$ the number of pixels). Their method generates an image $x$ by starting with random noise $y_0$ and iteratively taking steps along the gradient of the log prior by applying a denoiser. This approach allows the synthesis of realistic images that lie on the high-probability manifold of natural images.

The method can be extended to solve inverse linear problems by placing linear constraints on a noisy image through a low-rank measurement matrix \(
x^c = M^Tx \), where \(x^c\) is the vector of measurements. For example, in super-resolution, \(M^T\) represents a downsampling operation. Without loss of generality, Kadkhodaie and Simoncelli assume that \(M\) and \(M^T \) are semi-orthogonal, which implies that \(M\) is the
pseudo-inverse of \(M^T\) and that \(MM^T\)
can be used to project an image onto the measurement
subspace. With these constraints, the gradient of the conditional density can be derived as:
\begin{equation}
\sigma^2 \nabla_y \log{p(y|x^c)} = (I - MM^T)(\hat{x} - y) + M(x^c - M^Ty),
\end{equation}
where $I$ is the identity matrix. This conditional gradient can then be used within a gradient-ascent iterative scheme:
\begin{equation}
    y_{t} \leftarrow y_{t-1} + h_t \sigma^2 \nabla_y \log{p(y|x^c)} + \gamma_t z_t,
\end{equation}
where, at iteration $t$, the step size for the gradient is controlled by $h_t$ and the amount of noise $z_t \sim \mathcal{N}(0, \sigma^2=1)$ is controlled by $\gamma_t$. Scheduling for $h_t$ and $\gamma_t$ is parametrised by hyper-parameters $\sigma_0$, $\sigma_L$, $h_0$ and $\beta$ (see original paper). For convenience, the overall algorithm is reported in Algorithm~\ref{chp6:alg_simoncelli}.

\begin{algorithm}[t]
    \caption{Universal Inverse Solver as defined in~\cite{kadkhodaie_solving_2021}}
    \label{chp6:alg_simoncelli}
    \begin{algorithmic}[1]
        \State \textbf{Parameters:} $\sigma_0$, $\sigma_L$, $h_0$, $\beta$, ${ M}$, ${ x}^c$
        \State \textbf{Initialization:} $t \gets 1$; 
        \State Draw ${ y}_{0} \sim \mathcal{N}(0.5(I-{ M}{ M}^T){ e}+{ M}{ x}^c,\ \sigma_0^2 I)$, with ${ e}=[1, \dots, 1]^T$
        
        \While{$\sigma_{t-1} \leq \sigma_L$}
            \State $h_t \gets \frac{h_0 t}{1 + h_0 (t-1)}$
            \State ${ d}_t \gets (I-{ M}{ M}^T) ({\hat{x}}_{t-1} - {y}_{t-1}) + { M}({ x}^c - { M}^T { y}_{t-1})$
            \State $\gamma_t^2 \gets \left((1-\beta h_t)^2 -(1-h_t)^2\right) \frac{\| { d}_t \|^2}{N}$
            \State ${ y}_t \gets { y}_{t-1} + h_t { d}_t + \gamma_t { z}_t$ \ \ with ${ z}_t \sim \mathcal{N}(0,I)$
            \State $t \gets t + 1$
        \EndWhile
    \end{algorithmic}
\end{algorithm}

\begin{figure}[t]
\captionsetup{skip=0pt}
  \centering
  \includegraphics[width=\linewidth]{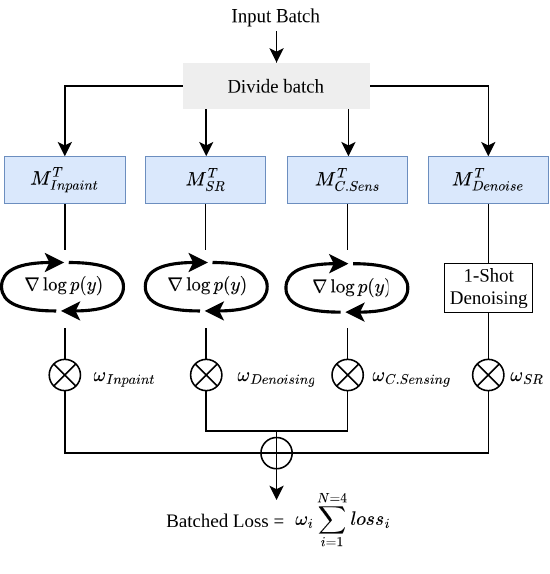}
  \caption{Illustration of our multi-task training framework. An input batch is divided into subsets for each inverse problem (inpainting, super-resolution, compressed sensing, and denoising). Each subset is passed through a task-specific forward operator \( M^T \) to produce the measurement vectors $x^c_i$, then, output images are iteratively reconstructed using a shared denoiser model that produces gradient estimates \( \nabla \log p(y) \). The circular arrows represent the iterative refinement process. Each task output is used to compute a reconstruction loss, weighted by task-specific coefficients \( \omega_i \), and summed to form the final training loss.}
  \Description{The diagram is a flowchart showing the multi-task training process for a denoiser. At the top, an input batch of images is split into four parallel branches, each representing a different task: inpainting, super-resolution, compressed sensing, and denoising. For the first three tasks, a light blue box applies a task-specific forward model, followed by an iterative loop that uses the denoiser to estimate the gradient of the image prior. The denoising branch instead performs a single-pass restoration without iteration.

Each branch outputs a reconstruction loss, which is then multiplied by a task-specific weight. These weighted losses are summed together at the bottom of the diagram to produce the batched loss used for training. This structure allows the same denoiser to be fine-tuned across multiple inverse problems simultaneously.
}

  \label{fig:single_model_training}
\end{figure}

\section{Fine-Tuning Denoisers For Solving Inverse Problems}

Kadkhodaie and Simoncelli's method hinges on using a pre-trained blind denoiser to approximate the log prior's gradient within their iterative solver. 
A key strength of this method is its adaptability, enabling the use of any compatible denoiser across various inverse problems. Recognising that the denoiser's characteristics will significantly affect the iterative method's outcome, a crucial question arises: can the denoiser be specifically optimised to enhance performance when applied in this manner to solve inverse problems?

\subsection{Differentiable Training Platform for the Denoiser}

To this end, we introduce a multi-task training platform designed to fine-tune a denoiser by integrating it directly into the iterative solver and training it on multiple linear inverse problem tasks (see Figure~\ref{fig:single_model_training}). A key aspect of our approach is treating the complete reconstruction as a differentiable computation graph. In training, a degraded input is fed through multiple iterations of the solver, with each step utilising the denoiser, until the final reconstruction is achieved. An image quality loss ($L_1$) is calculated on this final output and propagated backward through all intermediate stages to update the denoiser's parameters. Consequently, the network learns from a diverse distribution of intermediate reconstructions, enhancing its ability to generalize to the data encountered during iterative inverse problem solving.

\subsection{Fixing the Number of Iterations}
To enable training, we need to modify the original framework. 
In the original formulation, iterations proceed until the solution is deemed sufficiently smooth, specifically, when the standard deviation of the image residual falls below a predefined threshold. While appropriate for inference, this convergence-based stopping criterion is impractical for training, as it requires hundreds or thousands of iterations and results in inconsistent iteration counts across different images in the batch. To address this, we remove the smoothness-based termination condition and instead fix the number of iterations per training step. This not only standardises the training loop, but also enables backpropagation through a consistent unrolled graph. The selection of an appropriate iteration count and associated hyperparameters is addressed in Section~\ref{Sec:hyperparam_tuning}.

\subsection{Generalising the Denoiser via Multi-Task Training}
\label{sec:generalising_multiple_tasks}
To promote generalisation, we aim to fine-tune the denoiser not for a single inverse problem, but for a broader class of linear degradation processes. Rather than optimising it on a fixed task or constraint, we expose the denoiser to a variety of forward models during training. By applying the same network across different iterative reconstruction pipelines, the denoiser learns to produce gradient estimates that remain effective across a range of input conditions. This approach encourages the model to internalise a more flexible and broadly applicable image prior, rather than one tailored to a single inverse mapping. The specific set of tasks and their configuration are detailed in Section~\ref{Sec:chp6_results_multitask}.

Figure~\ref{fig:single_model_training} presents a generalised illustration of this training process. For each training step, a batch of \( B \) images is divided across \( N \) tasks. Each mini-batch is assigned a task and processed using a corresponding task-specific measurement operator \( M^\top_{\text{task}} \). The same denoiser is then applied iteratively within each task’s reconstruction pipeline to solve the constrained gradient ascent problem. This produces \( N \) task-specific losses, which are scaled by corresponding weighting factors (see Section~\ref{sec:loss_weighting}) and summed to produce the overall training loss used to update the denoiser parameters.



\section{Experimental Setup}

With the training framework outlined, we now need to establish the training and test datasets, the deep learning architecture used for fine-tuning during experimentation, as well as the hyperparameters that were tuned according to the fixed number of iterations. 


\subsection{Inverse Problem Tasks}\label{sec:chp6_measurement_matrices}

In our experiments, we include the tasks from the original paper, where the semi-orthogonal measurement matrices \(M^T\) project the image \(x\) onto a lower-dimensional subspace \(x^c\). We additionally introduce demosaicing as a new task.

\paragraph*{Super-Resolution}
\(M^T\) is a downsampling matrix represented as a convolution kernel $K$, where the convolution has stride 2 and kernel averages pixels over non-overlapping regions \( K = \frac{1}{4}\begin{bsmallmatrix}     1 & 1 \\ 1 & 1
\end{bsmallmatrix}\).

\paragraph*{Inpainting}
The measurement matrix \(M^T\) is a binary diagonal mask matrix, where the elements on the diagonal are set to zero or one, depending on whether that pixel falls within the missing region or not. In this task, the missing region is set to be \(16 \times 16\) block, randomly positioned.

\paragraph*{Compressive Sensing} This task~\cite{donoho2006compressed} aims to recover signals from a small number of random linear measurements. In this scenario, we will consider that only 10\% of the pixels are randomly preserved. This \(M^T\) matrix is defined in a similar way to the inpainting case.

\paragraph*{Demosaicing}
For demosaicing, \(M^T\) follows the Bayer Colour Filter Array (CFA) pattern, where each \(2 \times 2\) block contains one red, two green, and one blue pixel. The diagonal entries of \(M^T\) are, again, binary values indicating the availability of the colour channels.

\paragraph*{Frequency Super-Resolution}
In this process, the 2D FFT of \(x\) is computed, and a mask \(W\) is applied to zero out a fraction of the high-frequency coefficients (e.g., 20\%). Following the masking step, an inverse FFT is performed to obtain a low-resolution spatial-domain approximation:
\(x^c = \text{IFFT}(W \cdot \text{FFT}(x)),\)
where \(\text{FFT}(x)\) and \(\text{IFFT}\) represent the Fourier transform and its inverse, respectively.

\paragraph{Random Basis Projection}
For random basis projection, \(M^T\) maps the image vector \(x \in \mathbb{R}^{HW}\) onto a lower-dimensional subspace using a random semi-orthogonal measurement matrix \(M^T \in \mathbb{R}^{HW \times K}\), where \(K = HW / 5\).

\subsection{Training/Testing Set}  
The training dataset consists of 3,400 uncropped images selected from DIV2K~\cite{agustsson2017ntire} and LSDIR~\cite{li2023lsdir}. 

Our test set for all of our experiments will be the 100 images from the BSD100~\cite{martin2001database} dataset.

\subsection{Choice of Baseline Denoiser} 
Our choice of baseline denoiser follows, as in Kadkhodaie and Simoncelli~\cite{kadkhodaie_solving_2021}, a BF-CNN architecture. This is essentially a modification of DnCNN~\cite{mohan2019robust} that removes biases from convolutional and batch normalisation layers, enabling improved generalisation across a wide range of input noise levels.

\subsection{Hyperparameter Tuning for Fixed Number of Iterations}
\label{Sec:hyperparam_tuning}
\begin{table}[t]
\centering
\caption{Hyperparameters for Short and Long task configurations. The number of iterations per task is fixed to enable batched training, with other parameters (\(\beta\), \(h_0\), and \(\sigma_0\)) adjusted to ensure convergence.}
\begin{tabular}{@{}lllll@{}}
\toprule
Configuration & Iterations & $\beta$ & $h_0$ & $\sigma_0$ \\ \midrule
Short         & 25         & 0.4     & 0.10  & 0.6       \\
Medium          & 50         & 0.2     & 0.09  & 0.8        \\
Long          & 100         & 0.06     & 0.5  & 0.08        \\ \bottomrule
\end{tabular}%
\label{tab:hyperparams_configs}
\end{table}

To reduce the number of iterations, the smoothness criterion is replaced with a fixed iteration count based on task difficulty. By constraining the algorithm to a set number of iterations, hyperparameters must be adjusted to ensure convergence within this limit. The key hyperparameters are \(\beta\), which controls reintroduced noise; \(h_0\), the initial step size; and \(\sigma_0\), the starting noise level. For most tasks, starting with a heavily noised version of the constrained image (e.g., \(\sigma_0 = 0.8\)) is sufficient to achieve smooth results, avoiding the need for the higher noise levels (\(\sigma_0 = 1\)) typically used in the original framework.

We observed that different tasks converge at varying rates, which motivated the creation of separate iteration configurations: \textit{short}, \textit{medium}, and \textit{long}. Super-resolution and frequency super-resolution are assigned the \textit{short} configuration, while inpainting, compressed sensing, demosaicing, and random basis reconstruction are assigned the \textit{medium} configuration, as summarised in Table~\ref{tab:hyperparams_configs}. Additionally, a \textit{long} configuration is introduced to evaluate the potential benefits of allowing the solver to run for an extended number of iterations for all tasks.

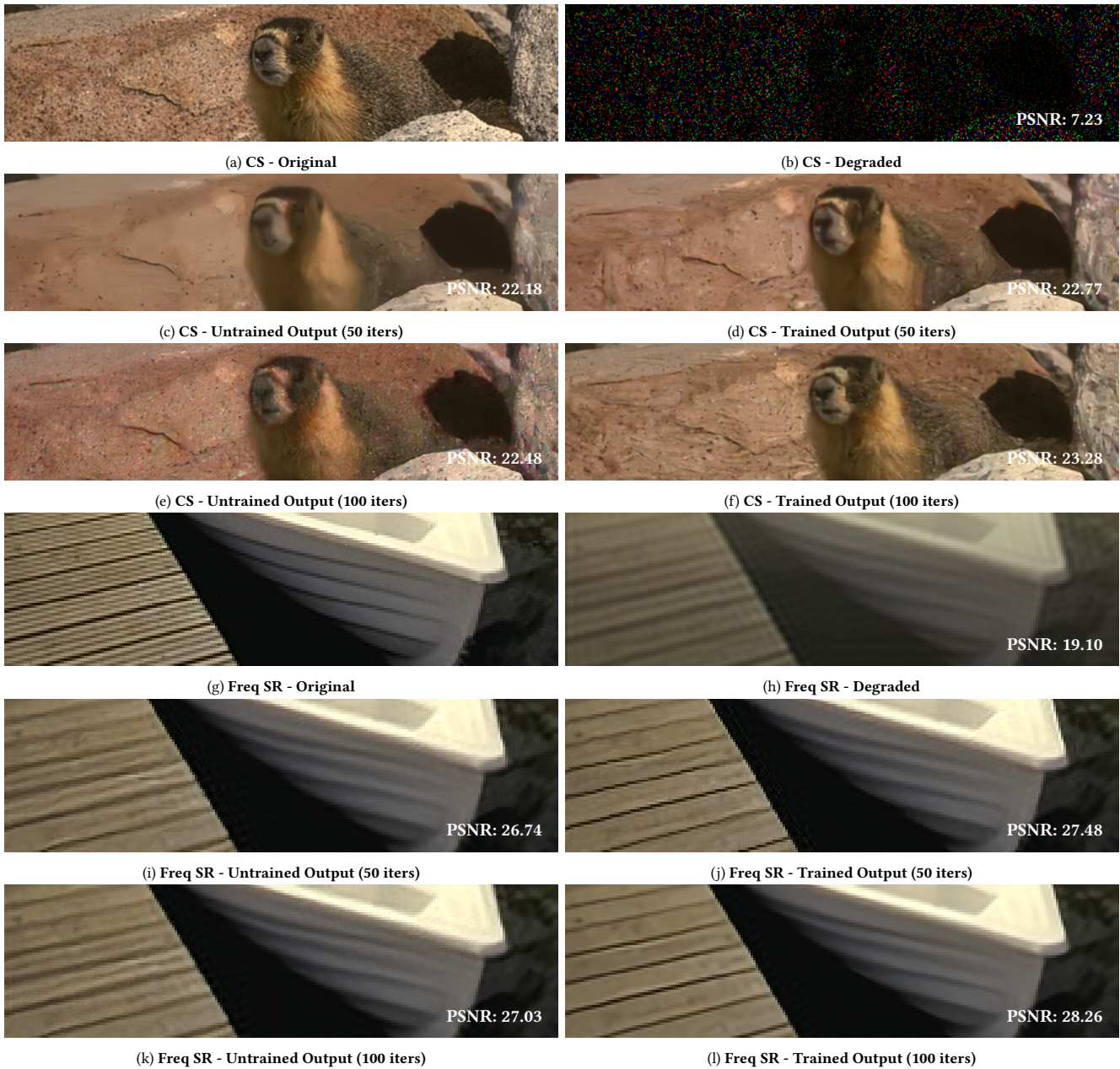
\begin{figure*}[t]
\captionsetup[sub]{font=footnotesize, labelfont=footnotesize}
\centering
\subcaptionbox{\footnotesize CS - Original\label{fig:cs_orig}}[0.49\linewidth]{
    \begin{tikzpicture}
        \node[inner sep=0pt] (img) at (0,0) {\includegraphics[width=\linewidth, trim=30 110 0 100, clip]{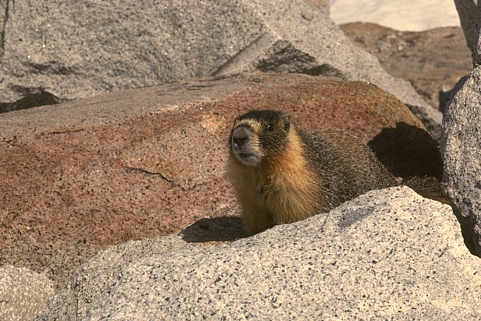}};
    \end{tikzpicture}
}%
\hspace{1pt}
\subcaptionbox{\footnotesize CS - Degraded\label{fig:cs_degraded}}[0.49\linewidth]{
    \begin{tikzpicture}
        \node[inner sep=0pt] (img) at (0,0) {\includegraphics[width=\linewidth, trim=30 110 0 100, clip]{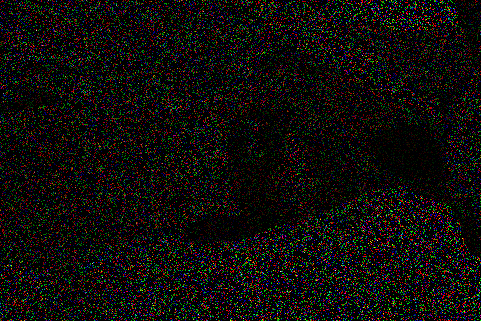}};
        \node[anchor=south east, text=white, font=\small, xshift=-4pt, yshift=4pt] at (img.south east) {\textbf{PSNR: 7.23}};
    \end{tikzpicture}
}
\subcaptionbox{\footnotesize CS - Untrained Output (50 iters)\label{fig:cs_untrained50}}[0.49\linewidth]{
    \begin{tikzpicture}
        \node[inner sep=0pt] (img) at (0,0) {\includegraphics[width=\linewidth, trim=30 110 0 100, clip]{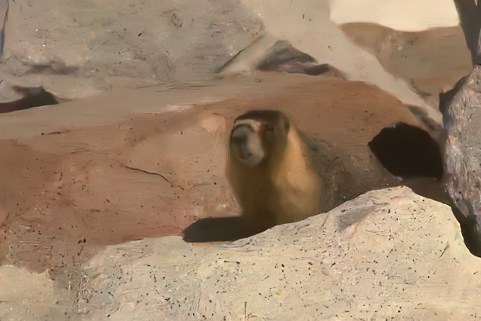}};
        \node[anchor=south east, text=white, font=\small, xshift=-4pt, yshift=4pt] at (img.south east){\textbf{PSNR: 22.18}};
    \end{tikzpicture}
}%
\hspace{1pt}
\subcaptionbox{\footnotesize CS - Trained Output (50 iters)\label{fig:cs_trained50}}[0.49\linewidth]{
    \begin{tikzpicture}
        \node[inner sep=0pt] (img) at (0,0) {\includegraphics[width=\linewidth, trim=30 110 0 100, clip]{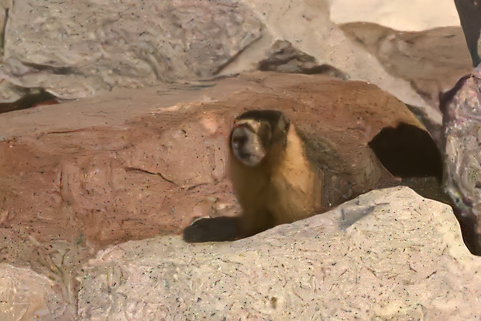}};
        \node[anchor=south east, text=white, font=\small, xshift=-4pt, yshift=4pt] at (img.south east) {\textbf{PSNR: 22.77}};
    \end{tikzpicture}
}
\subcaptionbox{\footnotesize CS - Untrained Output (100 iters)\label{fig:cs_untrained100}}[0.49\linewidth]{
    \begin{tikzpicture}
        \node[inner sep=0pt] (img) at (0,0) {\includegraphics[width=\linewidth, trim=30 110 0 100, clip]{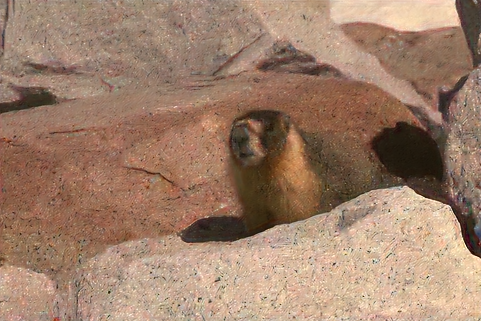}};
        \node[anchor=south east, text=white, font=\small, xshift=-4pt, yshift=4pt] at (img.south east) {\textbf{PSNR: 22.48}};
    \end{tikzpicture}
}%
\hspace{1pt}
\subcaptionbox{\footnotesize CS - Trained Output (100 iters)\label{fig:cs_trained100}}[0.49\linewidth]{
    \begin{tikzpicture}
        \node[inner sep=0pt] (img) at (0,0) {\includegraphics[width=\linewidth, trim=30 110 0 100, clip]{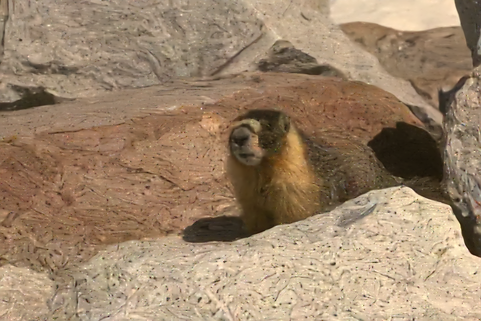}};
        \node[anchor=south east, text=white, font=\small, xshift=-4pt, yshift=4pt] at (img.south east) {\textbf{PSNR: 23.28}};
    \end{tikzpicture}
}

\subcaptionbox{\footnotesize Freq SR - Original\label{fig:fsr_orig}}[0.49\linewidth]{
    \begin{tikzpicture}
        \node[inner sep=0pt] (img) at (0,0) {\includegraphics[width=\linewidth, trim=0 20 100 400, clip]{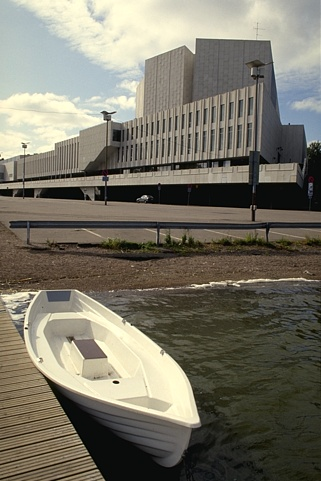}};
    \end{tikzpicture}
}%
\hspace{1pt}
\subcaptionbox{\footnotesize Freq SR - Degraded\label{fig:fsr_degraded}}[0.49\linewidth]{
    \begin{tikzpicture}
        \node[inner sep=0pt] (img) at (0,0) {\includegraphics[width=\linewidth, trim=0 20 100 400, clip]{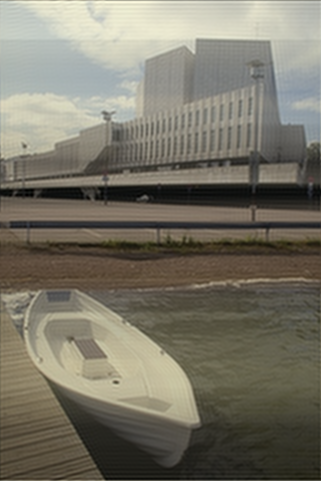}};
        \node[anchor=south east, text=white, font=\small, xshift=-4pt, yshift=4pt] at (img.south east) {\textbf{PSNR: 19.10}};
    \end{tikzpicture}
}
\subcaptionbox{\footnotesize Freq SR - Untrained Output (50 iters)\label{fig:fsr_untrained50}}[0.49\linewidth]{
    \begin{tikzpicture}
        \node[inner sep=0pt] (img) at (0,0) {\includegraphics[width=\linewidth, trim=0 20 100 400, clip]{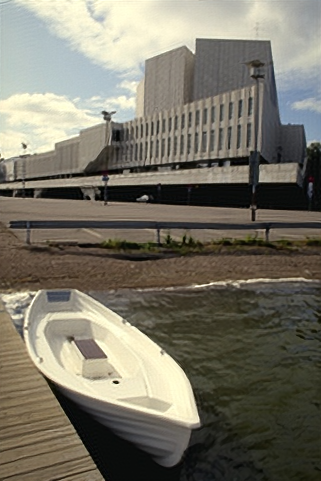}};
        \node[anchor=south east, text=white, font=\small, xshift=-4pt, yshift=4pt] at (img.south east) {\textbf{PSNR: 26.74}};
    \end{tikzpicture}
}%
\hspace{1pt}
\subcaptionbox{\footnotesize Freq SR - Trained Output (50 iters)\label{fig:fsr_trained50}}[0.49\linewidth]{
    \begin{tikzpicture}
        \node[inner sep=0pt] (img) at (0,0) {\includegraphics[width=\linewidth, trim=0 20 100 400, clip]{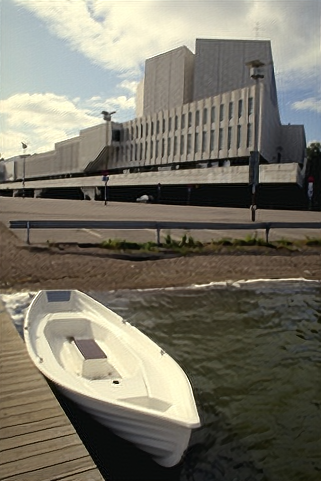}};
        \node[anchor=south east, text=white, font=\small, xshift=-4pt, yshift=4pt] at (img.south east) {\textbf{PSNR: 27.48}};
    \end{tikzpicture}
}
\subcaptionbox{\footnotesize Freq SR - Untrained Output (100 iters)\label{fig:fsr_untrained100}}[0.49\linewidth]{
    \begin{tikzpicture}
        \node[inner sep=0pt] (img) at (0,0) {\includegraphics[width=\linewidth, trim=0 20 100 400, clip]{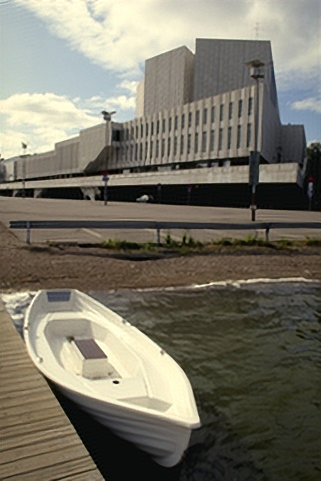}};
        \node[anchor=south east, text=white, font=\small, xshift=-4pt, yshift=4pt] at (img.south east) {\textbf{PSNR: 27.03}};
    \end{tikzpicture}
}%
\hspace{1pt}
\subcaptionbox{\footnotesize Freq SR - Trained Output (100 iters)\label{fig:fsr_trained100}}[0.49\linewidth]{
    \begin{tikzpicture}
        \node[inner sep=0pt] (img) at (0,0) {\includegraphics[width=\linewidth, trim=0 20 100 400, clip]{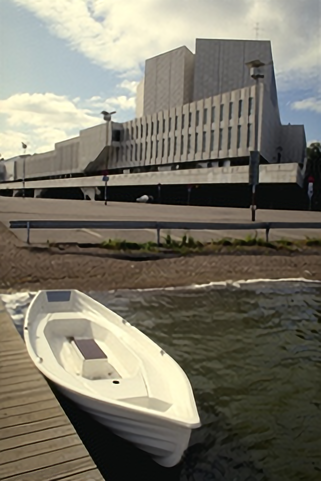}};
        \node[anchor=south east, text=white, font=\small, xshift=-4pt, yshift=4pt] at (img.south east) {\textbf{PSNR: 28.26}};
    \end{tikzpicture}
}
\caption{Sample results for two tasks: compressed sensing (top three rows) and frequency super-resolution (bottom three rows). PSNR values are shown in the bottom right of each image.}
\Description{The first set of comparisons shows compressive sensing restoration on an image of a marmot. There are six images altogether, the original image, the degraded image, the untrained denoiser restored image (50 iterations), the trained denoiser restored image (50 iterations), the untrained denoiser restored image (100 iterations), and the trained denoiser restored image (100 iterations) The second set of examples shows a close up of the front of a small rowboat and a wooden pier. This example is for frequency super-resolution. }
\label{fig:combined_cs_fsr}
\end{figure*}

\input{mosaic3_morecompressed}

\subsection{Training Details}

During training, input images are cropped to \(96 \times 96\) and augmented with random flips and rotations. Each task is supervised using the mean squared error (MSE) loss, and optimisation is performed with the AdamW optimiser over 500 epochs. A cosine annealing learning rate scheduler with warm restarts is employed: the learning rate is initially decreased from \(10^{-3}\) to zero over the first 100 epochs, then reset and decayed to zero every subsequent 50 epochs. To accommodate the high memory demands of unrolled backpropagation through the iterative solver, the batch size is limited to 16.

\section{Experiments}

In this section, we present a series of experiments designed to evaluate the feasibility and effectiveness of fine-tuning a denoiser within an iterative framework for linear inverse problems. We begin by assessing whether fine-tuning improves task performance compared to a fixed, pre-trained model. We then investigate the denoiser's ability to generalise to unseen tasks when trained jointly on multiple inverse problems. To address the challenge of balancing task-specific objectives during multi-task training, we explore various loss weighting strategies. Finally, we examine how fixing the number of iterations, rather than relying on convergence, impacts the quality and efficiency of the resulting reconstructions.


\subsection{Multi-Task Training Scenarios}\label{Sec:chp6_results_multitask}
We conduct a series of fine-tuning experiments to assess how well a denoiser can generalise across multiple inverse problems when trained with multiple objectives. Three tasks, frequency super-resolution, random basis reconstruction, and demosaicing, are excluded from training and instead used as benchmarks to evaluate generalisation to unseen problems.

All trained networks share the same DnCNN backbone and differ only in the set of tasks used during fine-tuning. We define each configuration as follows:

\begin{itemize}
    \item BF-CNN: The original, publicly available bias-free DnCNN model trained solely for denoising. This network is not fine-tuned further and serves as a fixed baseline.

    \item 3 Task: In this configuration, we fine-tune the BF-CNN model using three iterative tasks: inpainting, super-resolution, and compressed sensing. The denoising task is excluded.

    \item 3 Task + D: This configuration extends the 3 Task setup by including an additional denoising objective, where the noise standard deviation is sampled uniformly as \( \sigma \sim U(0, 100) \). Unlike the other tasks, denoising is performed in a single forward pass rather than iteratively. This reduces training complexity and reflects the structure of the reverse process in diffusion models.
\end{itemize}

\subsection{Loss Weighting Scenarios}
\label{sec:loss_weighting}
Some tasks produce loss values that converge to different scales due to varying task difficulties, causing tasks with smaller loss values to stop contributing to training early and underperform in testing, relative to their single-task training results. Essentially, the loss in the Multi-Task configuration amounts to a weighted average over the tasks   
  \(
    \mathcal{L}_{\mathrm{total}} = \sum^{N}_{i=0} w_i {\mathcal{L_{\mathrm{task_i}}}}
    \). 
We consider here to adjust the relative weights of each task according to different strategies:

\paragraph*{\textbf{Uniform}} In this configuration, each task is assigned an equal weight of 1: \( w_{\text{all}} = 1 \).
\paragraph*{\textbf{Fixed}} The model is biased towards the denoising tasks, with \( w_\text{denoise} = 5 \), while all other weights remain at 1.
\paragraph*{\textbf{Normalised}} The loss function is optimised to balance the impact of each task on gradient computation, preventing tasks with larger loss values from disproportionately influencing the optimisation process. The weight for each task is defined as \( w_i = 1 / \mathcal{L}^{\mathrm{prev}}_{\mathrm{task}_i} \), where \( \mathcal{L}^{\mathrm{prev}}_{\mathrm{task}_i} \) is the loss value of task \(i\) from a prior training run. For the denoising task, weights are interpolated based on the sampled noise standard deviation. The precomputed weights are summarised in Table~\ref{tab:chp6_lambda_Norm_values}.
\begin{table}[t]
\centering
\caption{Table of weights used for the equalised loss scheme in training. Here, \(\mathbf{w}\) are saved from previous losses to balance contributions from tasks.}
\resizebox{\linewidth}{!}{%
\begin{tabular}{@{}lccc cccccc@{}}
\toprule
     & Inpnt. & SRx2 & Sens. & \(\sigma = 5\) & \(\sigma = 10\) & \(\sigma = 20\) & \(\sigma = 30\) & \(\sigma = 40\) & \(\sigma = 50\) \\ \midrule
$\textbf{1/w}$ & 0.014 & 0.035 & 0.048 & 0.011 & 0.017 & 0.026 & 0.033 & 0.040 & 0.046 \\ 
\bottomrule
\end{tabular}%
}
\label{tab:chp6_lambda_Norm_values}
\end{table}

\subsection{Length of Iterations}
As discussed in Section~\ref{Sec:hyperparam_tuning}, we configure the framework to converge at a different number of iterations. In our experiments, we assess both the original BF-CNN model (with optimised scheduling parameters) and our fine-tuned model in the 50 iteration configuration and the 100 iteration configuration.

\input{main_results_table}
\section{Results}
\subsection{Linear Inverse Task Performance}
\input{final_noise_table}
In Table~\ref{tab:my-table_cleaned_with_deltas} we report PSNR results (in dB) for each network across both the trained tasks (inpainting, SR-\texttimes2, compressed sensing) and the untrained tasks (frequency SR, random basis reconstruction, demosaicing). Each value represents the mean result computed over 100 test images. For fine-tuned models, the difference in performance relative to the equivalent BF-CNN baseline is shown in parentheses. Across 12 experimental runs using different random seeds, the 95\% confidence intervals for PSNR remain within 0.09\,dB for all evaluated tasks, with the exception of demosaicing, which exhibits higher variability (±0.13\,dB). This indicates a high degree of consistency in the results across most tasks.

The \textit{3 Task\textsuperscript{U}} model, trained with a uniform loss weighting scheme for 50 iterations, yields substantial improvements on the tasks seen during training: inpainting improves by +4.08 dB, super-resolution by +1.66 dB, and compressed sensing by +2.09 dB. However, this model exhibits poor generalisation to the untrained tasks, most notably with a -6.90 dB drop on demosaicing compared to the baseline. As a result, it achieves the lowest overall mean improvement across all tasks.

When the denoising task is included during training (\textit{4 Task\textsuperscript{U}}), generalisation improves considerably, with a mean gain of +1.12 dB over the baseline. This inclusion also helps close the gap in demosaicing performance relative to the original denoiser.

Both the original and fine-tuned networks show improved performance when evaluated at 100 iterations. The best overall results are obtained by the \textit{4 Task\textsuperscript{N}} model, which achieves a mean PSNR of 32.53 dB—representing a +1.34 dB improvement over the baseline BF-CNN evaluated at the same iteration count.

\subsection{Denoising Performance}
In Table~\ref{tab:denoising_table}, we report one-shot denoising results for all iteratively trained models across a range of noise levels. As expected, the original BF-CNN model, trained exclusively for denoising, achieves the highest average PSNR (32.47 dB). In contrast, the \textit{3 Task\textsuperscript{U}} model, which excludes denoising during training, performs the worst, with an average PSNR drop of 4.65 dB relative to the baseline.

Introducing the denoising task into training substantially improves results. The \textit{4 Task\textsuperscript{U}} model recovers much of the lost performance, achieving an average within 0.93 dB of the original denoiser. Applying fixed loss weighting (\textit{4 Task\textsuperscript{F}}), which emphasises denoising, further narrows the gap to just 0.23 dB. Notably, the \textit{4 Task\textsuperscript{N}} model, trained with normalised loss weighting, performs nearly identically—just 0.22 dB below the baseline—while maintaining the strongest overall performance on the inverse problems. This indicates that normalised weighting provides a favourable trade-off between task-specific accuracy and broader generalisation.

Finally, the 100-iteration version of the same model (\textit{4 Task\textsuperscript{N}}) achieves slightly lower denoising performance (32.15 dB), suggesting a small degradation which may result from prolonged exposure to high noise levels during iterative training, or the distribution shift introduced by additional solver iterations.

\subsection{Visual Assessment}
Figure~\ref{fig:combined_cs_fsr} presents reconstructed samples from the test dataset for two tasks: compressive sensing (10\%) and frequency super-resolution. For each task, we show the original image, its degraded counterpart, the 50-iteration reconstruction using the original model (\textit{BF-CNN}), and the 100-iteration output using the fine-tuned model (\textit{4 Task\textsuperscript{N}}).

In the compressive sensing examples, notable differences are observed in texture synthesis. Across all samples, the iterative solver generates distinct textures in the rocky background. The trained models consistently recover sharper high-frequency details—especially around the rodent’s face and the surrounding terrain. For instance, the 100-iteration trained model is able to reconstruct fine hair on the rodent’s neck that is entirely absent from the BF-CNN output.

Similar improvements are observed in the frequency super-resolution samples. The wooden pier in the original image contains strong diagonal structures which are better preserved by the trained models. These structures appear blurred or softened in the BF-CNN output but are much more clearly defined after fine-tuning. Additionally, while both trained models improve edge fidelity, the 100-iteration variant produces a smoother result and reduces the staircasing artefacts that remain visible in the 50-iteration version.

Additional samples are provided in Figure~\ref{fig:final_grid_with_rows_4_5} comparing the baseline model, \textit{3 Task\textsuperscript{U} (50)}, \textit{4 Task\textsuperscript{N} (50)} and \textit{4 Task\textsuperscript{N} (100)}. 

\subsection{Summary of Findings}
Training a denoiser within the linear inverse solving framework proves effective, leading to consistent improvements across multiple restoration tasks. However, when denoising is excluded from the training process, denoising performance suffers significantly, highlighting its importance for generalisation. This is due to the large variations in loss magnitudes introduced by the denoising task during training, which complicate optimisation. Emphasising the denoising loss through a fixed weighting scheme improves denoising performance, however, the normalised loss scheme provides a more balanced result between denoising and linear inverse task accuracy. Additionally, extending the number of iterations during inference further enhances visual quality, producing smoother and more detailed reconstructions.

\section{Conclusions}
This work has shown that the denoiser responsible for generating the image prior gradient in an iterative linear inverse solving framework can be effectively fine-tuned to improve performance across multiple inverse problem tasks. By training a single lightweight denoising network jointly on a range of inverse problems, we achieve improved generalisation while reducing the number of required solver iterations compared to the original baseline. Our quantitative results demonstrate a mean improvement of 1.34~dB across all evaluated tasks, and qualitative assessments further reveal enhanced preservation of high-frequency detail and smoother reconstructions. These findings highlight the potential of jointly training denoisers within iterative frameworks, and open the door to future work involving larger, more expressive denoising networks.

\clearpage 
\bibliographystyle{ACM-Reference-Format}
\bibliography{refs,denoisingrefs}




\end{document}

%% file: mosaic3_morecompressed.tex
\newlength{\figmaxheight}
\setlength{\figmaxheight}{\dimexpr
  \textheight-\abovecaptionskip-\belowcaptionskip-2\baselineskip\relax}

\begin{figure*}[p]
    \captionsetup{aboveskip=0pt, belowskip=2pt} 
    \centering
    
    \begin{minipage}{\textwidth}
    \centering
        \begin{minipage}[t]{0.19\textwidth}
            \centering
            \textbf{\hspace{2mm}\scriptsize Degraded Projection $M^T(x_c)$}
        \end{minipage}%
        \begin{minipage}[t]{0.19\textwidth}
            \centering
            \textbf{\hspace{7mm} \scriptsize Original Denoiser (50)}
        \end{minipage}%
        \begin{minipage}[t]{0.19\textwidth}
            \centering
            \textbf{\hspace{7mm}\scriptsize 3 Task (50)}
        \end{minipage}%
        \begin{minipage}[t]{0.19\textwidth}
            \centering
            \textbf{\hspace{7mm}\scriptsize 4 Task (50)}
        \end{minipage}%
        \begin{minipage}[t]{0.19\textwidth}
            \centering
            \textbf{\hspace{7mm}\scriptsize 4 Task (100)} 
        \end{minipage}%
    \end{minipage}
    
    \begin{minipage}{\textwidth}
    \centering
        \begin{minipage}[t]{0.19\textwidth}
            \centering
            \begin{tikzpicture}[baseline, every node/.style={inner sep=0, outer sep=0}]
                \node[anchor=south west] at (0, 0) {\includegraphics[width=\textwidth, trim=10 30 10 10, clip]{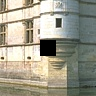}};
                \node[anchor=south, text=white] at (2.85, 0.1) {\footnotesize\texttt{\textbf{19.54 dB}}};
                \node[anchor=north west, rotate=90, text=white, fill=black, xshift=-21pt, yshift=-2pt] at (0, 1) {\LARGE{\texttt{\textbf{INPAINTING}}}};
            \end{tikzpicture}
        \end{minipage}%
        \begin{minipage}[t]{0.19\textwidth}
            \centering
            \begin{tikzpicture}[baseline, every node/.style={inner sep=0, outer sep=0}]
                \node[anchor=south west] at (0, 0) {\includegraphics[width=\textwidth, trim=10 30 10 10, clip]{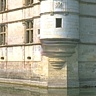}};
                \node[anchor=south, text=white] at (2.85, 0.1) {\footnotesize\texttt{\textbf{29.57 dB}}};
            \end{tikzpicture}
        \end{minipage}%
        \begin{minipage}[t]{0.19\textwidth}
            \centering
            \begin{tikzpicture}[baseline, every node/.style={inner sep=0, outer sep=0}]
                \node[anchor=south west] at (0, 0) {\includegraphics[width=\textwidth, trim=10 30 10 10, clip]{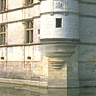}};
                \node[anchor=south, text=white] at (2.85, 0.1) {\footnotesize\texttt{\textbf{32.59 dB}}};
            \end{tikzpicture}
        \end{minipage}%
        \begin{minipage}[t]{0.19\textwidth}
            \centering
            \begin{tikzpicture}[baseline, every node/.style={inner sep=0, outer sep=0}]
                \node[anchor=south west] at (0, 0) {\includegraphics[width=\textwidth, trim=10 30 10 10, clip]{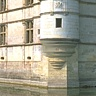}};
                \node[anchor=south, text=white] at (2.85, 0.1) {\footnotesize\texttt{\textbf{31.84 dB}}};
            \end{tikzpicture}
        \end{minipage}%
        \begin{minipage}[t]{0.19\textwidth}
            \centering
            \begin{tikzpicture}[baseline, every node/.style={inner sep=0, outer sep=0}]
                \node[anchor=south west] at (0, 0) {\includegraphics[width=\textwidth, trim=10 30 10 10, clip]{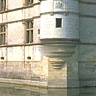}};
                \node[anchor=south, text=white] at (2.85, 0.1) {\footnotesize{\textbf{31.35 dB}}};
            \end{tikzpicture}
        \end{minipage}%
    \end{minipage}

    \begin{minipage}{\textwidth}
    \centering
        \begin{minipage}[t]{0.19\textwidth}
            \centering
            \begin{tikzpicture}[baseline, every node/.style={inner sep=0, outer sep=0}]
                \node[anchor=south west] at (0, 0) {\includegraphics[width=\textwidth, trim=95 60 95 40, clip]{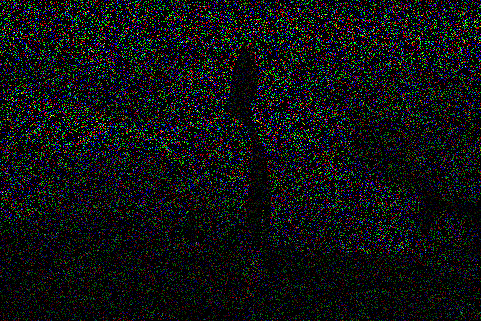}};
                \node[anchor=south, text=white] at (2.85, 0.1) {\footnotesize\texttt{\textbf{3.41 dB}}};
                \node[anchor=north west, rotate=90, text=white, xshift=-20pt, yshift=-2pt] at (0, 1) {\LARGE{\texttt{\textbf{C. SENSING}}}};
            \end{tikzpicture}
        \end{minipage}%
        \begin{minipage}[t]{0.19\textwidth}
            \centering
            \begin{tikzpicture}[baseline, every node/.style={inner sep=0, outer sep=0}]
                \node[anchor=south west] at (0, 0) {\includegraphics[width=\textwidth, trim=95 60 95 40, clip]{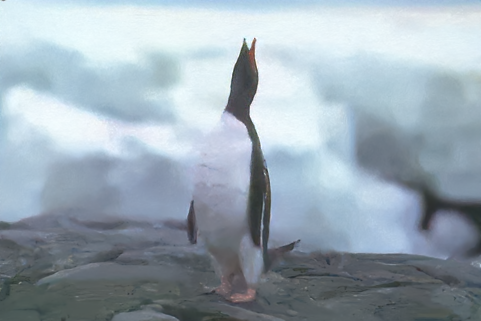}};
                \node[anchor=south, text=white] at (2.85, 0.1) {\footnotesize\texttt{\textbf{29.63 dB}}};
            \end{tikzpicture}
        \end{minipage}%
        \begin{minipage}[t]{0.19\textwidth}
            \centering
            \begin{tikzpicture}[baseline, every node/.style={inner sep=0, outer sep=0}]
                \node[anchor=south west] at (0, 0) {\includegraphics[width=\textwidth, trim=95 60 95 40, clip]{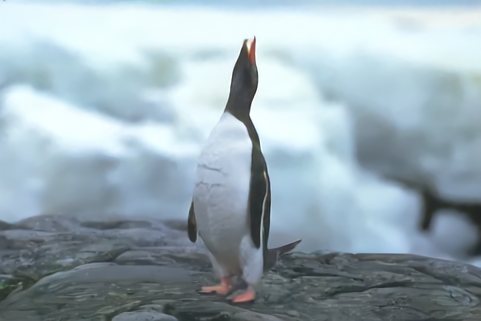}};
                \node[anchor=south, text=white] at (2.85, 0.1) {\footnotesize\texttt{\textbf{32.56 dB}}};
            \end{tikzpicture}
        \end{minipage}%
        \begin{minipage}[t]{0.19\textwidth}
            \centering
            \begin{tikzpicture}[baseline, every node/.style={inner sep=0, outer sep=0}]
                \node[anchor=south west] at (0, 0) {\includegraphics[width=\textwidth, trim=95 60 95 40, clip]{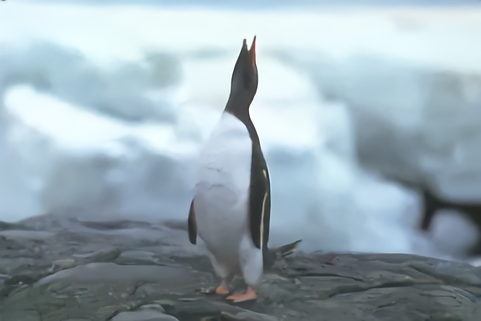}};
                \node[anchor=south, text=white] at (2.85, 0.1) {\footnotesize\texttt{\textbf{31.75 dB}}};
            \end{tikzpicture}
        \end{minipage}%
        \begin{minipage}[t]{0.19\textwidth}
            \centering
            \begin{tikzpicture}[baseline, every node/.style={inner sep=0, outer sep=0}]
                \node[anchor=south west] at (0, 0) {\includegraphics[width=\textwidth, trim=95 60 95 40, clip]{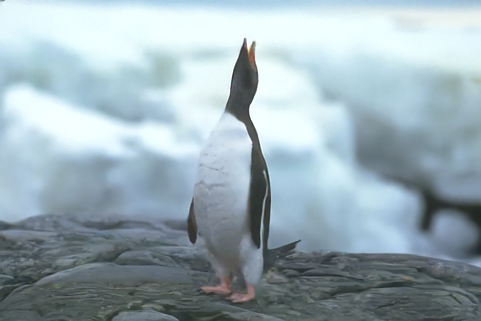}};
                \node[anchor=south, text=white] at (2.85, 0.1) {\footnotesize\texttt{\textbf{33.74 dB}}};
            \end{tikzpicture}
        \end{minipage}%
    \end{minipage}

    \begin{minipage}{\textwidth}
    \centering
        \begin{minipage}[t]{0.19\textwidth}
            \centering
            \begin{tikzpicture}[baseline, every node/.style={inner sep=0, outer sep=0}]
                \node[anchor=south west] at (0, 0) {\includegraphics[width=\textwidth, trim=120 90 95 40, clip]{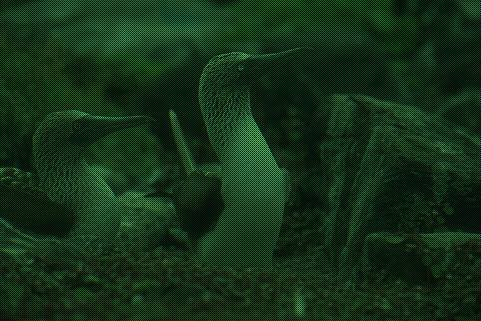}};
                \node[anchor=south, text=white] at (2.85, 0.1) {\footnotesize\texttt{\textbf{10.21 dB}}};
                \node[anchor=north west, rotate=90, text=white, xshift=-15pt, yshift=-2pt] at (0, 1) {\LARGE{\texttt{\textbf{DEMOSAIC}}}};
            \end{tikzpicture}
        \end{minipage}%
        \begin{minipage}[t]{0.19\textwidth}
            \centering
            \begin{tikzpicture}[baseline, every node/.style={inner sep=0, outer sep=0}]
                \node[anchor=south west] at (0, 0) {\includegraphics[width=\textwidth, trim=120 90 95 40, clip]{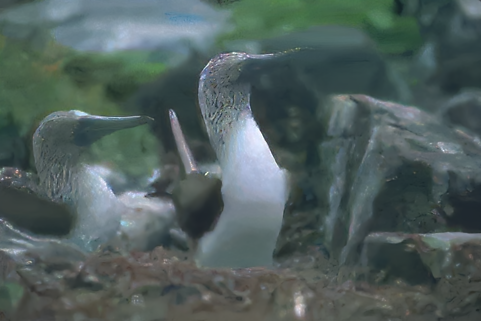}};
                \node[anchor=south, text=white] at (2.85, 0.1) {\footnotesize\texttt{\textbf{27.33 dB}}};
            \end{tikzpicture}
        \end{minipage}%
        \begin{minipage}[t]{0.19\textwidth}
            \centering
            \begin{tikzpicture}[baseline, every node/.style={inner sep=0, outer sep=0}]
                \node[anchor=south west] at (0, 0) {\includegraphics[width=\textwidth, trim=120 90 95 40, clip]{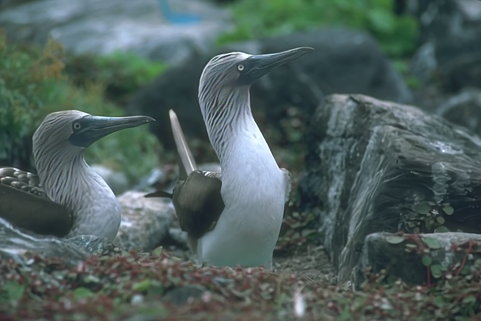}};
                \node[anchor=south, text=white] at (2.85, 0.1) {\footnotesize\texttt{\textbf{35.68 dB}}};
            \end{tikzpicture}
        \end{minipage}%
        \begin{minipage}[t]{0.19\textwidth}
            \centering
            \begin{tikzpicture}[baseline, every node/.style={inner sep=0, outer sep=0}]
                \node[anchor=south west] at (0, 0) {\includegraphics[width=\textwidth, trim=120 90 95 40, clip]{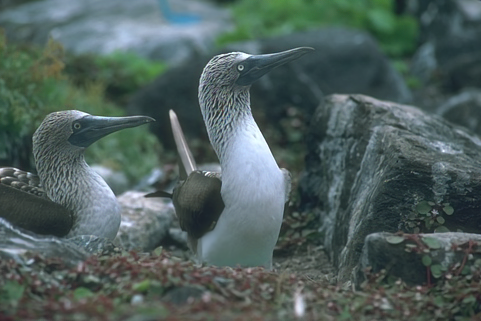}};
                \node[anchor=south, text=white] at (2.85, 0.1) {\footnotesize\texttt{\textbf{41.70 dB}}};
            \end{tikzpicture}
        \end{minipage}%
        \begin{minipage}[t]{0.19\textwidth}
            \centering
            \begin{tikzpicture}[baseline, every node/.style={inner sep=0, outer sep=0}]
                \node[anchor=south west] at (0, 0) {\includegraphics[width=\textwidth, trim=120 90 95 40, clip]{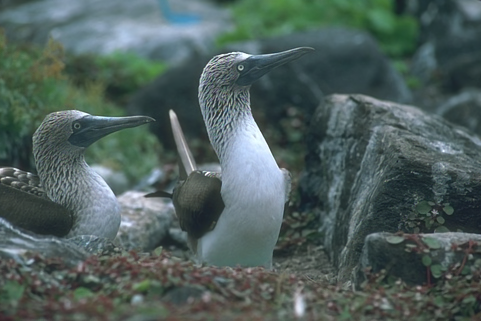}};
                \node[anchor=south, text=white] at (2.85, 0.1) {\footnotesize\texttt{\textbf{42.83 dB}}};
            \end{tikzpicture}
        \end{minipage}%
    \end{minipage}
    
    \begin{minipage}{\textwidth}
    \centering
        \begin{minipage}[t]{0.19\textwidth}
            \centering
            \begin{tikzpicture}[baseline, every node/.style={inner sep=0, outer sep=0}]
                \node[anchor=south west] at (0, 0) {\includegraphics[width=\textwidth, trim=0 20 0 0, clip]{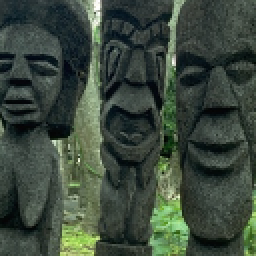}};
                \node[anchor=south, text=white] at (2.85, 0.1) {\footnotesize\texttt{\textbf{24.05 dB}}};
                \node[anchor=north west, rotate=90, text=white, xshift=3pt, yshift=-2pt] at (0, 1) {\LARGE{\texttt{\textbf{X2 SR}}}};
            \end{tikzpicture}
        \end{minipage}%
        \begin{minipage}[t]{0.19\textwidth}
            \centering
            \begin{tikzpicture}[baseline, every node/.style={inner sep=0, outer sep=0}]
                \node[anchor=south west] at (0, 0) {\includegraphics[width=\textwidth, trim=0 20 0 0, clip]{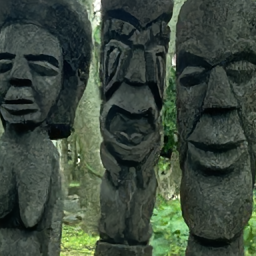}};
                \node[anchor=south, text=white] at (2.85, 0.1) {\footnotesize\texttt{\textbf{24.21 dB}}};
            \end{tikzpicture}
        \end{minipage}%
        \begin{minipage}[t]{0.19\textwidth}
            \centering
            \begin{tikzpicture}[baseline, every node/.style={inner sep=0, outer sep=0}]
                \node[anchor=south west] at (0, 0) {\includegraphics[width=\textwidth, trim=0 20 0 0, clip]{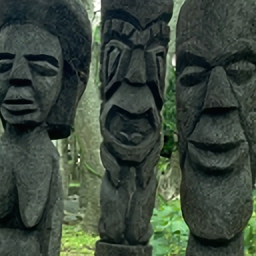}};
                \node[anchor=south, text=white] at (2.85, 0.1) {\footnotesize\texttt{\textbf{25.14 dB}}};
            \end{tikzpicture}
        \end{minipage}%
        \begin{minipage}[t]{0.19\textwidth}
            \centering
            \begin{tikzpicture}[baseline, every node/.style={inner sep=0, outer sep=0}]
                \node[anchor=south west] at (0, 0) {\includegraphics[width=\textwidth, trim=0 20 0 0, clip]{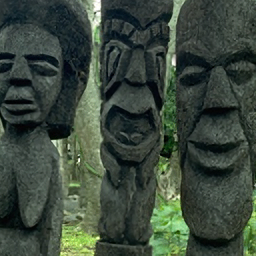}};
                \node[anchor=south, text=white] at (2.85, 0.1) {\footnotesize\texttt{\textbf{24.45 dB}}};
            \end{tikzpicture}
        \end{minipage}%
        \begin{minipage}[t]{0.19\textwidth}
            \centering
            \begin{tikzpicture}[baseline, every node/.style={inner sep=0, outer sep=0}]
                \node[anchor=south west] at (0, 0) {\includegraphics[width=\textwidth, trim=0 20 0 0, clip]{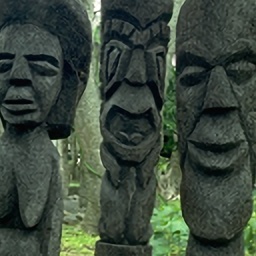}};
                \node[anchor=south, text=white] at (2.85, 0.1) {\footnotesize\texttt{\textbf{25.17 dB}}};
            \end{tikzpicture}
        \end{minipage}%
    \end{minipage}

    \begin{minipage}{\textwidth}
    \centering
        \begin{minipage}[t]{0.19\textwidth}
            \centering
            \begin{tikzpicture}[baseline, every node/.style={inner sep=0, outer sep=0}]
                \node[anchor=south west] at (0, 0) {\includegraphics[width=\textwidth, trim=15 40 0 20, clip]{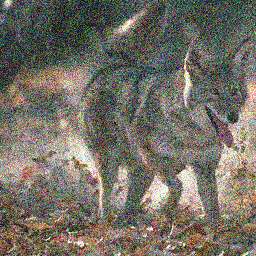}};
                \node[anchor=south, text=white] at (2.85, 0.1) {\footnotesize\texttt{\textbf{16.11 dB}}};
                \node[anchor=north west, rotate=90, text=white, xshift=-15pt, yshift=-2pt] at (0, 1) {\LARGE{\texttt{\textbf{DENOISING}}}};
            \end{tikzpicture}
        \end{minipage}%
        \begin{minipage}[t]{0.19\textwidth}
            \centering
            \begin{tikzpicture}[baseline, every node/.style={inner sep=0, outer sep=0}]
                \node[anchor=south west] at (0, 0) {\includegraphics[width=\textwidth, trim=15 40 0 20, clip]{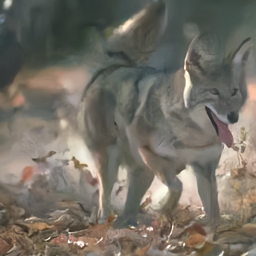}};
                \node[anchor=south, text=white] at (2.85, 0.1) {\footnotesize\texttt{\textbf{29.60 dB}}};
            \end{tikzpicture}
        \end{minipage}%
        \begin{minipage}[t]{0.19\textwidth}
            \centering
            \begin{tikzpicture}[baseline, every node/.style={inner sep=0, outer sep=0}]
                \node[anchor=south west] at (0, 0) {\includegraphics[width=\textwidth, trim=15 40 0 20, clip]{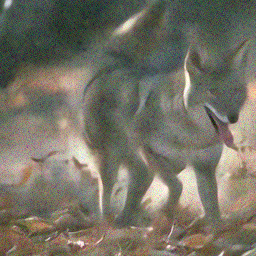}};
                \node[anchor=south, text=white] at (2.85, 0.1) {\footnotesize\texttt{\textbf{23.11 dB}}};
            \end{tikzpicture}
        \end{minipage}%
        \begin{minipage}[t]{0.19\textwidth}
            \centering
            \begin{tikzpicture}[baseline, every node/.style={inner sep=0, outer sep=0}]
                \node[anchor=south west] at (0, 0) {\includegraphics[width=\textwidth, trim=15 40 0 20, clip]{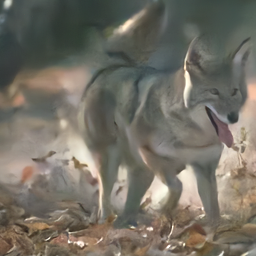}};
                \node[anchor=south, text=white] at (2.85, 0.1) {\footnotesize\texttt{\textbf{29.46 dB}}};
            \end{tikzpicture}
        \end{minipage}%
        \begin{minipage}[t]{0.19\textwidth}
            \centering
            \begin{tikzpicture}[baseline, every node/.style={inner sep=0, outer sep=0}]
                \node[anchor=south west] at (0, 0) {\includegraphics[width=\textwidth, trim=15 40 0 20, clip]{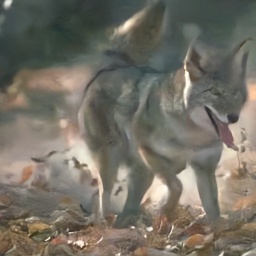}};
                \node[anchor=south, text=white] at (2.85, 0.1) {\footnotesize\texttt{\textbf{29.51 dB}}};
            \end{tikzpicture}
        \end{minipage}%
    \end{minipage}

    \begin{minipage}{\textwidth}
    \centering
        \begin{minipage}[t]{0.19\textwidth}
            \centering
            \begin{tikzpicture}[baseline, every node/.style={inner sep=0, outer sep=0}]
                \node[anchor=south west] at (0, 0) {\includegraphics[width=\textwidth, trim=0 5 0 10, clip]{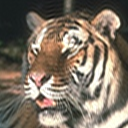}};
                \node[anchor=south, text=white] at (2.85, 0.1) {\footnotesize\texttt{\textbf{30.55 dB}}};
                \node[anchor=north west, rotate=90, text=white, xshift=-5pt, yshift=-2pt] at (0, 1) {\LARGE{\texttt{\textbf{FREQ SR}}}};
            \end{tikzpicture}
        \end{minipage}%
        \begin{minipage}[t]{0.19\textwidth}
            \centering
            \begin{tikzpicture}[baseline, every node/.style={inner sep=0, outer sep=0}]
                \node[anchor=south west] at (0, 0) {\includegraphics[width=\textwidth, trim=0 5 0 10, clip]{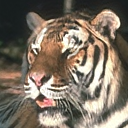}};
                \node[anchor=south, text=white] at (2.85, 0.1) {\footnotesize\texttt{\textbf{30.88 dB}}};
            \end{tikzpicture}
        \end{minipage}%
        \begin{minipage}[t]{0.19\textwidth}
            \centering
            \begin{tikzpicture}[baseline, every node/.style={inner sep=0, outer sep=0}]
                \node[anchor=south west] at (0, 0) {\includegraphics[width=\textwidth, trim=0 5 0 10, clip]{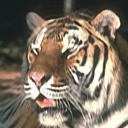}};
                \node[anchor=south, text=white] at (2.85, 0.1) {\footnotesize\texttt{\textbf{31.85 dB}}};
            \end{tikzpicture}
        \end{minipage}%
        \begin{minipage}[t]{0.19\textwidth}
            \centering
            \begin{tikzpicture}[baseline, every node/.style={inner sep=0, outer sep=0}]
                \node[anchor=south west] at (0, 0) {\includegraphics[width=\textwidth, trim=0 5 0 10, clip]{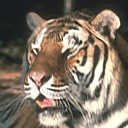}};
                \node[anchor=south, text=white] at (2.85, 0.1) {\footnotesize\texttt{\textbf{31.44 dB}}};
            \end{tikzpicture}
        \end{minipage}%
        \begin{minipage}[t]{0.19\textwidth}
            \centering
            \begin{tikzpicture}[baseline, every node/.style={inner sep=0, outer sep=0}]
                \node[anchor=south west] at (0, 0) {\includegraphics[width=\textwidth, trim=0 5 0 10, clip]{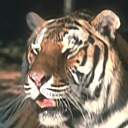}};
                \node[anchor=south, text=white] at (2.85, 0.1) {\footnotesize\texttt{\textbf{32.05 dB}}};
            \end{tikzpicture}
        \end{minipage}%
    \end{minipage}

    \begin{minipage}{\textwidth}
    \centering
        \begin{minipage}[t]{0.19\textwidth}
            \centering
            \begin{tikzpicture}[baseline, every node/.style={inner sep=0, outer sep=0}]
                \node[anchor=south west] at (0, 0) {\includegraphics[width=\textwidth, trim=0 10 0 10, clip]{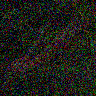}};
                \node[anchor=south, text=white] at (2.85, 0.1) {\footnotesize\texttt{\textbf{8.56 dB}}};
                \node[anchor=north west, rotate=90, text=white, xshift=-18pt, yshift=-2pt] at (0, 1) {\LARGE{\texttt{\textbf{RAND BASIS}}}};
            \end{tikzpicture}
        \end{minipage}%
        \begin{minipage}[t]{0.19\textwidth}
            \centering
            \begin{tikzpicture}[baseline, every node/.style={inner sep=0, outer sep=0}]
                \node[anchor=south west] at (0, 0) {\includegraphics[width=\textwidth, trim=0 10 0 10, clip]{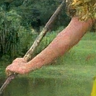}};
                \node[anchor=south, text=white] at (2.85, 0.1) {\footnotesize\texttt{\textbf{27.94 dB}}};
            \end{tikzpicture}
        \end{minipage}%
        \begin{minipage}[t]{0.19\textwidth}
            \centering
            \begin{tikzpicture}[baseline, every node/.style={inner sep=0, outer sep=0}]
                \node[anchor=south west] at (0, 0) {\includegraphics[width=\textwidth, trim=0 10 0 10, clip]{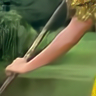}};
                \node[anchor=south, text=white] at (2.85, 0.1) {\footnotesize\texttt{\textbf{26.77 dB}}};
            \end{tikzpicture}
        \end{minipage}%
        \begin{minipage}[t]{0.19\textwidth}
            \centering
            \begin{tikzpicture}[baseline, every node/.style={inner sep=0, outer sep=0}]
                \node[anchor=south west] at (0, 0) {\includegraphics[width=\textwidth, trim=0 10 0 10, clip]{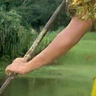}};
                \node[anchor=south, text=white] at (2.85, 0.1) {\footnotesize\texttt{\textbf{28.39 dB}}};
            \end{tikzpicture}
        \end{minipage}%
        \begin{minipage}[t]{0.19\textwidth}
            \centering
            \begin{tikzpicture}[baseline, every node/.style={inner sep=0, outer sep=0}]
                \node[anchor=south west] at (0, 0) {\includegraphics[width=\textwidth, trim=0 10 0 10, clip]{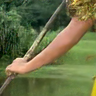}};
                \node[anchor=south, text=white] at (2.85, 0.1) {\footnotesize\texttt{\textbf{29.51 dB}}};
            \end{tikzpicture}
        \end{minipage}%
    \end{minipage}
    \caption{Sample Results of trained denoisers for inverse solving. Each column represents a different denoiser and each row represents a different task. From left to right we show the projection of the constrained measurement matrix $x_c$, the result of using the original BF-CNN denoiser prior, the 3 task fine-tuned prior and the 4-task, normalised loss fine-tuned denoiser. From top to bottom, the following tasks are shown: inpainting, \texttimes2 super-resolution, $\sigma=40$ AWGN denoising, frequency super resolution, 10\% random basis reconstruction.}
    \label{fig:final_grid_with_rows_4_5}
    \Description{A grid of examples of our trained inverse solver in various tasks. Each row shows a new image  and a new task. From top to bottom the rows show, a section of a castle for inpainting, a penguin for compressive sensing restoration, tribal statues for spatial super resolution, a wolf for denoising, a crop of a tiger's head for frequency super resolution, and a hand holding an oar with a river and grass in the background for random basis restoration. For each row, there are 5 columns: the degraded input, the original untrained denoiser (50 iterations), the 3 task trained denoiser (50 iterations), the 4 trask trained denoiser (50 iterations) and the 4 task trained denoiser (100 iterations).}
\end{figure*}

%% file: main_results_table.tex
\begin{table*}[t]
\centering
\caption{PSNR Results (dB) for solving linear inverse problems using various denoiser training strategies. \textbf{BF-CNN} denotes the model weights used by Kadkhodaie, \textbf{3 Task} is trained on three iterative tasks: inpainting, $\times$2 super-resolution, and compressed sensing. \textbf{4 Task} adds a fourth, non-iterative denoising task ($\sigma \sim U(0,100)$). Superscripts indicate the loss weighting scheme: \textsuperscript{U} = uniform, \textsuperscript{F} = fixed, \textsuperscript{N} = normalised. Differences (in parentheses) are relative to BF-CNN at the same iteration setting. With the exception of demosaicing, all values lie within 0.09\,dB 95\% confidence intervals across repeated runs.}
\begin{tblr}{
    colspec={Q[c]Q[l,t]Q[l,m]Q[c]Q[c]Q[c]Q[c]Q[c]Q[c]Q[c]Q[c]},
    width=\linewidth,
}
\toprule[2pt]
& & Degraded & BF-CNN & 3 Task\textsuperscript{U} & 4 Task\textsuperscript{U} & 4 Task\textsuperscript{F} & 4 Task\textsuperscript{N} & BF-CNN & 4 Task\textsuperscript{N} & \\ 
&Iterations & - & 50 & 50 & 50 & 50 & 50 & 100 & 100 & \\ 
\midrule
\SetCell[r=3]{c}\rotatebox{90}{\textit{trained}} & Inpainting     & 23.27 & 33.67 & \textbf{37.75} (+4.08) & 36.89 (+3.22) & 36.61 (+2.94) & 37.36 (+3.69) & 36.16 & 37.41 (+1.25) & \\
& SR-\texttimes2 & 27.32 & 28.06 & 29.72 (+1.66) & 29.22 (+1.16) & 28.61 (+0.55) & 27.36 (-0.70) & 28.78 & \textbf{29.78} (+1.00) & \\
& Sensing        & 7.33  & 24.79 & 26.88 (+2.09) & 26.28 (+1.49) & 25.54 (+0.75) & 26.27 (+1.48) & 25.22 & \textbf{27.09} (+1.87) & \\
\hline[dashed]
\SetCell[r=3]{c}\rotatebox{90}{\textit{untrained}} & Freq. SR       & 27.65 & 29.92 & 30.95 (+1.03) & 30.59 (+0.67) & 30.24 (+0.32) & 29.38 (-0.54) & 30.30 & \textbf{31.23} (+0.93) & \\ 
& Rand. Basis    & 7.53  & 29.26 & 27.66 (-1.60) & 30.46 (+1.20) & 29.36 (+0.10) & 30.50 (+1.24) & 29.10 & \textbf{31.45} (+2.35) & \\
& Demosaic       & 8.68  & \textbf{38.74} & 31.84 (-6.90) & 37.70 (-1.04) & 36.36 (-2.38) & 37.47 (-1.27) & 37.59 & 38.22 (+0.63) & \\
\midrule
& Mean           & 16.96 & 30.74 & 30.80 (+0.06) & 31.86 (+1.12) & 31.12 (+0.38) & 31.39 (+0.65) & 31.19 & \textbf{32.53} (+1.34) & \\
\bottomrule[2pt]
\end{tblr}

\label{tab:my-table_cleaned_with_deltas}
\end{table*}

%% file: final_noise_table.tex
\begin{table*}[t]
\centering
\caption{PSNR (dB) Denoising results across varying noise levels $\sigma$.}
\begin{tblr}{
    colspec={Q[l,t]Q[l,m]Q[r]Q[r]Q[r]Q[r]Q[r]Q[r]Q[r]},
    width=\linewidth,
    row{3} = {rowsep=0mm},
    rows = {rowsep=0mm},
    column{1} = {rightsep=1pt},
    column{2} = {leftsep=1pt, rightsep=3pt},
    column{3} = {leftsep=4pt},
}
\toprule[2pt]
\SetCell[r=1,c=2]{f,l}Model & & {Input} & {BF-CNN} & {3-Task\textsuperscript{U}} & {4-Task\textsuperscript{U}} & {4-Task\textsuperscript{F}} & {4-Task\textsuperscript{N}} & {4-Task\textsuperscript{N}} \\ 
\SetCell[r=1,c=2]{f,l}Iterations & & 50 & 50 & 50 & 50 & 50 & 50 & 100 \\ \midrule
\SetCell[r=6]{t}{\rotatebox{90}{\textit{Noise STD.}}} 
& $\sigma=5$   & 34.17 & \textbf{39.94} & 37.50 (-2.44) & 39.19 (-0.75) & 39.20 (-0.74) & 39.60 (-0.34) & 39.25 (-0.35) \\
& $\sigma=10$  & 28.15 & \textbf{35.91} & 32.95 (-2.96) & 35.56 (-0.35) & 35.71 (-0.20) & 35.79 (-0.12) & 35.35 (-0.44) \\
& $\sigma=20$  & 22.13 & \textbf{32.17} & 27.92 (-4.25) & 31.59 (-0.58) & 32.09 (-0.08) & 32.01 (-0.16) & 31.92 (-0.25) \\
& $\sigma=30$  & 18.61 & \textbf{30.16} & 24.84 (-5.32) & 29.29 (-0.87) & 30.07 (-0.09) & 29.97 (-0.19) & 29.99 (-0.31) \\
& $\sigma=40$  & 16.11 & \textbf{28.81} & 22.67 (-6.14) & 27.56 (-1.25) & 28.70 (-0.11) & 28.58 (-0.23) & 28.67 (-0.13) \\
& $\sigma=50$  & 14.17 & \textbf{27.82} & 21.01 (-6.81) & 26.04 (-1.78) & 27.68 (-0.14) & 27.52 (-0.30) & 27.68 (-0.14) \\
\midrule
\SetCell[r=1,c=2]{}Mean & & 22.22 & \textbf{32.47} & 27.82 (-4.65) & 31.54 (-0.93) & 32.24 (-0.23) & 32.25 (-0.22) & 32.15 (-0.32) \\
\bottomrule[2pt]
\end{tblr}
\label{tab:denoising_table}
\end{table*}

%% file: denoisingrefs.bib
@inproceedings{guo2019toward,
  title={Toward convolutional blind denoising of real photographs},
  author={Guo, Shi and Yan, Zifei and Zhang, Kai and Zuo, Wangmeng and Zhang, Lei},
  booktitle={Proceedings of the IEEE/CVF conference on computer vision and pattern recognition},
  pages={1712--1722},
  year={2019}
}

@inproceedings{liu2021swin,
  title={Swin transformer: Hierarchical vision transformer using shifted windows},
  author={Liu, Ze and Lin, Yutong and Cao, Yue and Hu, Han and Wei, Yixuan and Zhang, Zheng and Lin, Stephen and Guo, Baining},
  booktitle={Proceedings of the IEEE/CVF International Conference on Computer Vision},
  pages={10012--10022},
  year={2021}
}

@article{zhang_beyond_2017,
  title={Beyond a gaussian denoiser: Residual learning of deep cnn for image denoising},
  author={Zhang, Kai and Zuo, Wangmeng and Chen, Yunjin and Meng, Deyu and Zhang, Lei},
  journal={IEEE transactions on image processing},
  volume={26},
  number={7},
  pages={3142--3155},
  year={2017},
  publisher={IEEE}
}

@inproceedings{tassano2019dvdnet,
  title={Dvdnet: A fast network for deep video denoising},
  author={Tassano, Matias and Delon, Julie and Veit, Thomas},
  booktitle={2019 IEEE International Conference on Image Processing (ICIP)},
  pages={1805--1809},
  year={2019},
  organization={IEEE}
}

@inproceedings{tassano2020fastdvdnet,
  title={Fastdvdnet: Towards real-time deep video denoising without flow estimation},
  author={Tassano, Matias and Delon, Julie and Veit, Thomas},
  booktitle={Proceedings of the IEEE/CVF conference on computer vision and pattern recognition},
  pages={1354--1363},
  year={2020}
}

@article{liang2022vrt,
author = {Liang, Jingyun and Cao, Jiezhang and Fan, Yuchen and Zhang, Kai and Ranjan, Rakesh and Li, Yawei and Timofte, Radu and Van Gool, Luc},
title = {VRT: A Video Restoration Transformer},
year = {2024},
issue_date = {2024},
publisher = {IEEE Press},
volume = {33},
issn = {1057-7149},
url = {https://doi.org/10.1109/TIP.2024.3372454},
doi = {10.1109/TIP.2024.3372454},
journal = {Trans. Img. Proc.},
month = jan,
pages = {2171–2182},
numpages = {12}
}

@article{yue2023rvideformer,
  title={RViDeformer: Efficient Raw Video Denoising Transformer with a Larger Benchmark Dataset},
  author={Yue, Huanjing and Cao, Cong and Liao, Lei and Yang, Jingyu},
  journal={arXiv e-prints},
  pages={arXiv--2305},
  year={2023}
}

@inproceedings{qiao2017learning,
  title={Learning non-local image diffusion for image denoising},
  author={Qiao, Peng and Dou, Yong and Feng, Wensen and Li, Rongchun and Chen, Yunjin},
  booktitle={Proceedings of the 25th ACM international conference on Multimedia},
  pages={1847--1855},
  year={2017}
}

@inproceedings{nichol2021improved,
  title={Improved denoising diffusion probabilistic models},
  author={Nichol, Alexander Quinn and Dhariwal, Prafulla},
  booktitle={International conference on machine learning},
  pages={8162--8171},
  year={2021},
  organization={PMLR}
}

@article{yan2020depth,
  title={Depth image denoising using nuclear norm and learning graph model},
  author={Yan, Chenggang and Li, Zhisheng and Zhang, Yongbing and Liu, Yutao and Ji, Xiangyang and Zhang, Yongdong},
  journal={ACM Transactions on Multimedia Computing, Communications, and Applications (TOMM)},
  volume={16},
  number={4},
  pages={1--17},
  year={2020},
  publisher={ACM New York, NY, USA}
}

@inproceedings{ren2022robust,
  title={Robust low-rank convolution network for image denoising},
  author={Ren, Jiahuan and Zhang, Zhao and Hong, Richang and Xu, Mingliang and Zhang, Haijun and Zhao, Mingbo and Wang, Meng},
  booktitle={Proceedings of the 30th ACM International Conference on Multimedia},
  pages={6211--6219},
  year={2022}
}

@inproceedings{bled2022assessing,
  title={Assessing advances in real noise image denoisers},
  author={Bled, Clement and Pitie, Francois},
  booktitle={Proceedings of the 19th ACM SIGGRAPH European Conference on Visual Media Production},
  pages={1--9},
  year={2022}
}


%% file: refs.bib
@article{miyasawa1961empirical,
  title={An empirical Bayes estimator of the mean of a normal population},
  author={Miyasawa, Koichi and others},
  journal={Bull. Inst. Internat. Statist},
  volume={38},
  number={181-188},
  pages={1--2},
  year={1961}
}

@article{kadkhodaie2021stochastic,
  title={Stochastic solutions for linear inverse problems using the prior implicit in a denoiser},
  author={Kadkhodaie, Zahra and Simoncelli, Eero},
  journal={Advances in Neural Information Processing Systems},
  volume={34},
  pages={13242--13254},
  year={2021}
}

@article{song2020denoising,
  title={Denoising diffusion implicit models},
  author={Song, Jiaming and Meng, Chenlin and Ermon, Stefano},
  journal={arXiv preprint arXiv:2010.02502},
  year={2020}
}

@article{dhariwal2021diffusion,
  title={Diffusion models beat gans on image synthesis},
  author={Dhariwal, Prafulla and Nichol, Alexander},
  journal={Advances in neural information processing systems},
  volume={34},
  pages={8780--8794},
  year={2021}
}

@article{kadkhodaie_solving_2021,
  title={Solving linear inverse problems using the prior implicit in a denoiser},
  author={Kadkhodaie, Zahra and Simoncelli, Eero P},
  journal={arXiv preprint arXiv:2007.13640},
  year={2020}
}

@inproceedings{sohl2015deep,
  title={Deep unsupervised learning using nonequilibrium thermodynamics},
  author={Sohl-Dickstein, Jascha and Weiss, Eric and Maheswaranathan, Niru and Ganguli, Surya},
  booktitle={International conference on machine learning},
  pages={2256--2265},
  year={2015},
  organization={PMLR}
}

@inproceedings{venkatakrishnan2013plug,
  title={Plug-and-play priors for model based reconstruction},
  author={Venkatakrishnan, Singanallur V and Bouman, Charles A and Wohlberg, Brendt},
  booktitle={2013 IEEE global conference on signal and information processing},
  pages={945--948},
  year={2013},
  organization={IEEE}
}

@article{boyd2011distributed,
  title={Distributed optimization and statistical learning via the alternating direction method of multipliers},
  author={Boyd, Stephen and Parikh, Neal and Chu, Eric and Peleato, Borja and Eckstein, Jonathan and others},
  journal={Foundations and Trends{\textregistered} in Machine learning},
  volume={3},
  number={1},
  pages={1--122},
  year={2011},
  publisher={Now Publishers, Inc.}
}

@inproceedings{brifman2016turning,
  title={Turning a denoiser into a super-resolver using plug and play priors},
  author={Brifman, Alon and Romano, Yaniv and Elad, Michael},
  booktitle={2016 IEEE International Conference on Image Processing (ICIP)},
  pages={1404--1408},
  year={2016},
  organization={IEEE}
}

@article{romano2017little,
  title={The little engine that could: Regularization by denoising (RED)},
  author={Romano, Yaniv and Elad, Michael and Milanfar, Peyman},
  journal={SIAM Journal on Imaging Sciences},
  volume={10},
  number={4},
  pages={1804--1844},
  year={2017},
  publisher={SIAM}
}

@inproceedings{zhang_learning_2017,
  title     = {Learning {Deep} {CNN} {Denoiser} {Prior} for {Image} {Restoration}},
  author    = {Zhang, Kai and Zuo, Wangmeng and Gu, Shuhang and Zhang, Lei},
  booktitle = {Proceedings of the IEEE Conference on Computer Vision and Pattern Recognition (CVPR)},
  year      = {2017},
  pages     = {3929--3938},
  url       = {https://openaccess.thecvf.com/content_cvpr_2017/html/Zhang_Learning_Deep_CNN_CVPR_2017_paper.html},
  urldate   = {2024-06-21},
}

@article{zhang2021plug,
  title={Plug-and-play image restoration with deep denoiser prior},
  author={Zhang, Kai and Li, Yawei and Zuo, Wangmeng and Zhang, Lei and Van Gool, Luc and Timofte, Radu},
  journal={IEEE Transactions on Pattern Analysis and Machine Intelligence},
  volume={44},
  number={10},
  pages={6360--6376},
  year={2021},
  publisher={IEEE}
}

@incollection{robbins1992empirical,
  title={An empirical Bayes approach to statistics},
  author={Robbins, Herbert E},
  booktitle={Breakthroughs in Statistics: Foundations and basic theory},
  pages={388--394},
  year={1992},
  publisher={Springer}
}

@article{efron2011tweedie,
  title={Tweedie’s formula and selection bias},
  author={Efron, Bradley},
  journal={Journal of the American Statistical Association},
  volume={106},
  number={496},
  pages={1602--1614},
  year={2011},
  publisher={Taylor \& Francis}
}

@article{mohan2019robust,
  title={Robust and interpretable blind image denoising via bias-free convolutional neural networks},
  author={Mohan, Sreyas and Kadkhodaie, Zahra and Simoncelli, Eero P and Fernandez-Granda, Carlos},
  journal={arXiv preprint arXiv:1906.05478},
  year={2019}
}

@inproceedings{agustsson2017ntire,
  title={Ntire 2017 challenge on single image super-resolution: Dataset and study},
  author={Agustsson, Eirikur and Timofte, Radu},
  booktitle={Proceedings of the IEEE conference on computer vision and pattern recognition workshops},
  pages={126--135},
  year={2017}
}

@inproceedings{li2023lsdir,
  title={Lsdir: A large scale dataset for image restoration},
  author={Li, Yawei and Zhang, Kai and Liang, Jingyun and Cao, Jiezhang and Liu, Ce and Gong, Rui and Zhang, Yulun and Tang, Hao and Liu, Yun and Demandolx, Denis and others},
  booktitle={Proceedings of the IEEE/CVF Conference on Computer Vision and Pattern Recognition},
  pages={1775--1787},
  year={2023}
}

@inproceedings{martin2001database,
  title={A database of human segmented natural images and its application to evaluating segmentation algorithms and measuring ecological statistics},
  author={Martin, David and Fowlkes, Charless and Tal, Doron and Malik, Jitendra},
  booktitle={Proceedings eighth IEEE international conference on computer vision. ICCV 2001},
  volume={2},
  pages={416--423},
  year={2001},
  organization={IEEE}
}

@article{donoho2006compressed,
  title={Compressed sensing},
  author={Donoho, David L},
  journal={IEEE Transactions on information theory},
  volume={52},
  number={4},
  pages={1289--1306},
  year={2006},
  publisher={IEEE}
}

@article{daras2024survey,
  title={A survey on diffusion models for inverse problems},
  author={Daras, Giannis and Chung, Hyungjin and Lai, Chieh-Hsin and Mitsufuji, Yuki and Ye, Jong Chul and Milanfar, Peyman and Dimakis, Alexandros G and Delbracio, Mauricio},
  journal={arXiv preprint arXiv:2410.00083},
  year={2024}
}

@inproceedings{wu2024jores,
  title={JoReS-Diff: Joint Retinex and Semantic Priors in Diffusion Model for Low-light Image Enhancement},
  author={Wu, Yuhui and Wang, Guoqing and Wang, Zhiwen and Yang, Yang and Li, Tianyu and Zhang, Malu and Li, Chongyi and Shen, Heng Tao},
  booktitle={Proceedings of the 32nd ACM International Conference on Multimedia},
  pages={1810--1818},
  year={2024}
}

@inproceedings{wu2024diffusion,
  title={Diffusion Posterior Proximal Sampling for Image Restoration},
  author={Wu, Hongjie and He, Linchao and Zhang, Mingqin and Chen, Dongdong and Luo, Kunming and Luo, Mengting and Zhou, Ji-Zhe and Chen, Hu and Lv, Jiancheng},
  booktitle={Proceedings of the 32nd ACM International Conference on Multimedia},
  pages={214--223},
  year={2024}
}

@inproceedings{lu20243d,
  title={3d priors-guided diffusion for blind face restoration},
  author={Lu, Xiaobin and Hu, Xiaobin and Luo, Jun and Zhu, Ben and Ruan, Yaping and Ren, Wenqi},
  booktitle={Proceedings of the 32nd ACM International Conference on Multimedia},
  pages={1829--1838},
  year={2024}
}

@inproceedings{tan2024blind,
  title={Blind face video restoration with temporal consistent generative prior and degradation-aware prompt},
  author={Tan, Jingfan and Park, Hyunhee and Zhang, Ying and Wang, Tao and Zhang, Kaihao and Kong, Xiangyu and Dai, Pengwen and Liu, Zikun and Luo, Wenhan},
  booktitle={Proceedings of the 32nd ACM International Conference on Multimedia},
  pages={1417--1426},
  year={2024}
}
